\documentclass[11pt]{article}
\usepackage{moriond,epsfig}

\def\be{\begin{equation}}
\def\ee{\end{equation}}
\def\bea{\begin{eqnarray}}
\def\eea{\end{eqnarray}}

\def\to{\rightarrow}

\let\<=\langle
\let\>=\rangle

\def\beq{\begin{equation}}
\def\eeq{\end{equation}}
\def\ba{\begin{array}}
\def\bea{\begin{eqnarray}}
\def\ea{\end{array}}
\def\eea{\end{eqnarray}}
\def\bit{\begin{itemize}}
\def\eit{\end{itemize}}

\def\ee#1{ \times 10^{#1} }
\def\comment#1{ \hbox{[{\it Comment suppressed here.}\/]} }
\def\hide#1{}

\newcommand{\nn}{\nonumber }

\def\Vud{0.9744(5)(3)}
\def\Vus{~~~0.225(2)(1)~~~}
\def\Vub{3.5(5)(5)\!\times\! 10^{-3}}
\def\Vcd{~~~0.24(3)(2)~~~}
\def\Vcs{~~~~0.97(10)(2)~~~~}
\def\Vcb{3.9(1)(3)\!\times\! 10^{-2}}
\def\Vtd{8.1(2.7)\!\times\! 10^{-3}}
\def\Vts{3.8(4)(3)\!\times\! 10^{-2}}
\def\Vtb{0.9992(0)(1)}
\def\Wlambda{\Vus}
\def\WA{0.77(2)(7)}
\def\WR{0.40(6)(6)}
\def\Wrho{0.16(28)}
\def\Weta{0.36(11)}
\begin{document}
\title{$B,~D,~K$ decays and CKM matrix from lattice QCD}

\author{Masataka Okamoto
}

\address{Fermi National Accelerator Laboratory, 
Batavia, Illinois 60510, USA
}

\maketitle\abstracts{
We use lattice QCD to fully determine the CKM matrix.
$|V_{cd}|$, $|V_{cs}|$, $|V_{ub}|$, $|V_{cb}|$ and $|V_{us}|$
are, respectively, directly determined
with recent lattice results 
for form factors of
semileptonic $D\to \pi l\nu$,
             $D\to K l\nu$,
             $B\to \pi l\nu$,
             $B\to D l\nu$
         and $K\to \pi l\nu$ decays
obtained by the Fermilab Lattice, MILC, and HPQCD Collaborations.
In addition,
$|V_{ud}|$, $|V_{tb}|$, $|V_{ts}|$ and $|V_{td}|$ are determined
by using unitarity of the CKM matrix and the experimental result
for $\sin{(2\beta)}$.
}

\section{Introduction}

The Cabibbo-Kobayashi-Maskawa (CKM) matrix,
which relates the mass eigenstates and the weak eigenstates
in the Standard Model electroweak theory, is a set of parameters.
To determine each CKM matrix element, one requires both
theoretical and experimental inputs.
On the theoretical side, one needs to know relevant hadronic
amplitudes, which often contain nonperturbative QCD effects.
A major role of lattice QCD is to calculate such hadronic
amplitudes reliably and accurately, from first principles.
One can then extract the CKM matrix elements by combining 
lattice QCD as the theoretical input
with the experimental input such as decay rates.
In this paper, we show that it is now possible to {\it fully} determine
the CKM matrix, for the first time, using lattice QCD.
The result for the full CKM matrix with lattice QCD is:
        \bea V_{\rm CKM} ~=~
        \left(
        \begin{array}{ccc}
        {{|V_{ud}|}}   & {|V_{us}|}   & {|V_{ub}|} \\
            \Vud       &    \Vus      &    \Vub    \\
        { |V_{cd}| }   & {|V_{cs}|}   & {|V_{cb}|} \\
            \Vcd       &    \Vcs      &    \Vcb    \\
        { |V_{td}| }   & {|V_{ts}|}   & {|V_{tb}|} \\
            \Vtd       &    \Vts      &    \Vtb    \\
        \end{array}
        \right) \label{ckm}
        \eea
where the first errors are from lattice calculations
and the second are experimental, except the one for 
$|V_{td}|$ which is a combined lattice and experimental error.
The results for the Wolfenstein parameters with lattice QCD are:
\bea
\lambda = \!\!\!\!\Wlambda\!\!\!\!\!\!,~~~~
A = \WA ,~~~~
\rho = \Wrho ,~~~~
\eta = \Weta.
\label{Wolf}
\eea

To directly determine 5 CKM matrix elements 
($|V_{cd}|$, $|V_{cs}|$, $|V_{ub}|$, $|V_{cb}|$ and $|V_{us}|$),
we use 5 semileptonic decays 
($D\to \pi l\nu$,
 $D\to K l\nu$,
 $B\to \pi l\nu$,
 $B\to D l\nu$ and
 $K\to \pi l\nu$),
for which the techniques for lattice calculations are well
established, and thus reliable calculations are possible.
The accuracy of previous lattice calculations
was limited
by two large systematic uncertainties ---
the error from the ``quenched'' approximation (neglect of
virtual quark loop effects)
and the error from the ``chiral'' 
extrapolation in light quark mass ($m_l\to m_{ud}$).
Both led to effects of around 10--20\%.
Recent work by the Fermilab Lattice, MILC, and HPQCD 
Collaborations~\cite{Aubin:2004ej,Okamoto:2004xg} 
successfully reduces these two dominant uncertainties.
The error from the quenched approximation is removed by
using the MILC unquenched gauge configurations~\cite{milc},
where the effect of $u,d$ and $s$ quarks is included ($n_f=2+1$).
The error from the chiral extrapolation is greatly 
reduced by using improved staggered quarks. 
With this improved approach, the accuracy of the 5 CKM matrix elements
is comparable to that of the Particle Data Group~\cite{Eidelman:wy}. 
The results for $|V_{ub}|$, $|V_{cb}|$ and $|V_{us}|$ are preliminary.
%
We then use CKM unitarity
to determine the other
4 CKM matrix elements ($|V_{ud}|$, $|V_{tb}|$, $|V_{ts}|$ and $|V_{td}|$).
In this way, we obtain all 9 CKM matrix elements 
and all the Wolfenstein parameters.
%
The results for the $D$ and $B$ decays have been presented in 
Refs.~\cite{Aubin:2004ej,Okamoto:2004xg}.
This work is a part of ongoing project 
of flavor physics with lattice QCD
by the Fermilab Lattice, MILC, and HPQCD Collaborations;
see also
\cite{Simone:2004fr,diPierro:2003iw,Okamoto:2004df}.

\section{5 CKM matrix elements from 5 semileptonic decays}
\label{sec:SLdecay}

\subsection{$D\to\pi(K)l\nu$, $|V_{cd(s)}|$ and $B\to\pi l\nu$, $|V_{ub}|$}
The differential decay rate $d\Gamma/dq^2$ for 
the heavy-to-light semileptonic decay $H\to Pl\nu$
is proportional to $|V_{ij}|^2 |f_+(q^2)|^2$,
where $f_+$ is a form factor of the relevant hadronic amplitude
defined through
\bea
\< P | V^\mu | H \>
 &=& 
f_+(q^2) (p_H+p_P-\Delta)^\mu + f_0(q^2) \Delta^\mu \label{eq:HLff}\\
&=& 
\sqrt{2m_H} \, \left[v^\mu \, f_\parallel(E) +
p^\mu_\perp \, f_\perp(E) \right]. \nn
\eea
Here $q = p_H - p_P$, $\Delta^\mu=(m_H^2-m_P^2)\, q^\mu / q^2$,
$v=p_H/m_H$, $p_\perp=p_P-Ev$ and $E=E_P$.
To determine the CKM matrix element $|V_{ij}|$ with the
experimental rate
$\int^{q^2_{\rm max}}_{q^2_{\rm min}}dq^2\ (d\Gamma/dq^2)$,
we calculate $f_{+,0}$ as a function of $q^2$.
Below we briefly describe the analysis 
procedure in Refs.~\cite{Aubin:2004ej,Okamoto:2004xg}.

\begin{figure}[tb]
\begin{center}
\centerline{
\epsfig{file=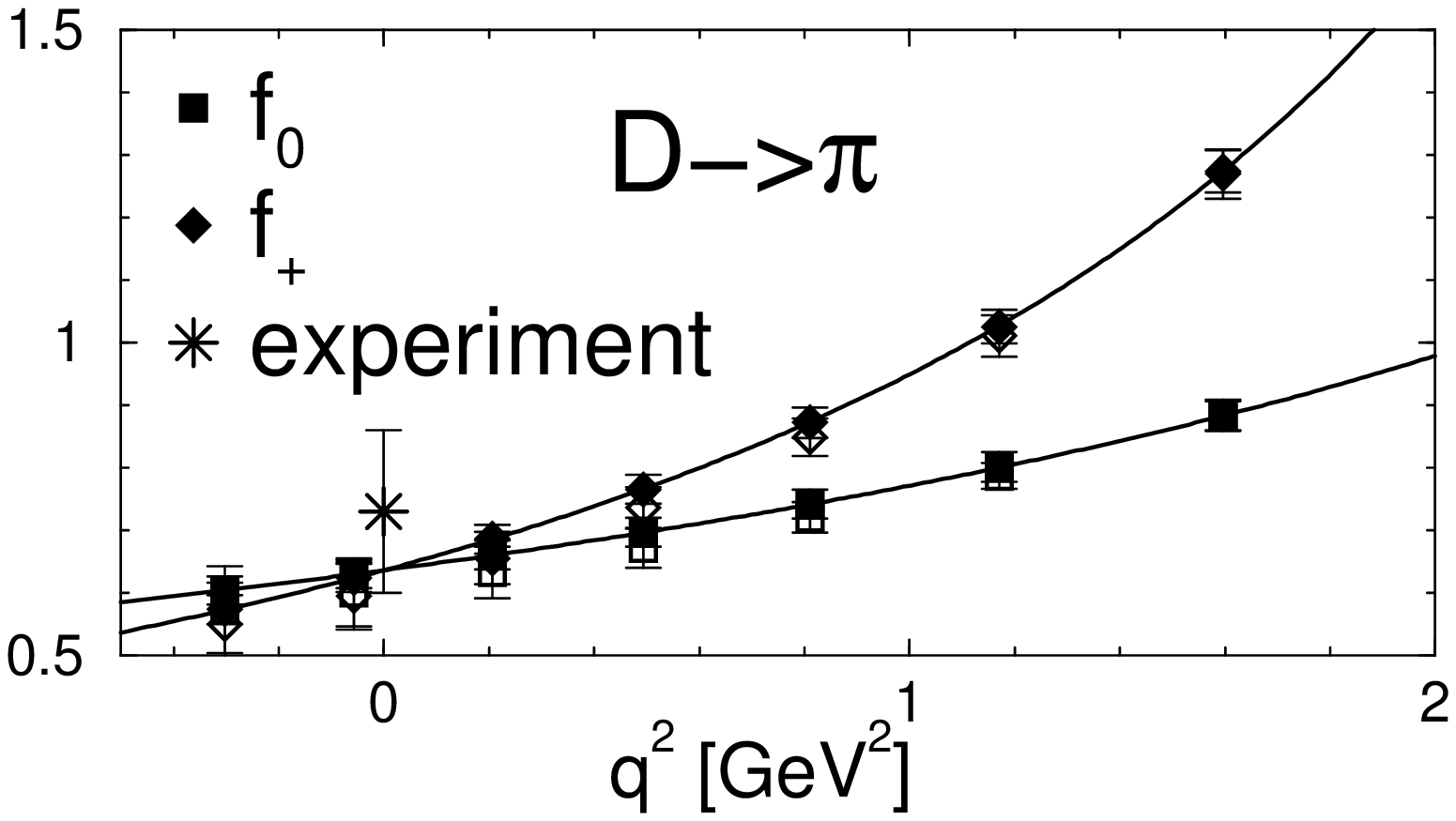,height=37mm}
\epsfig{file=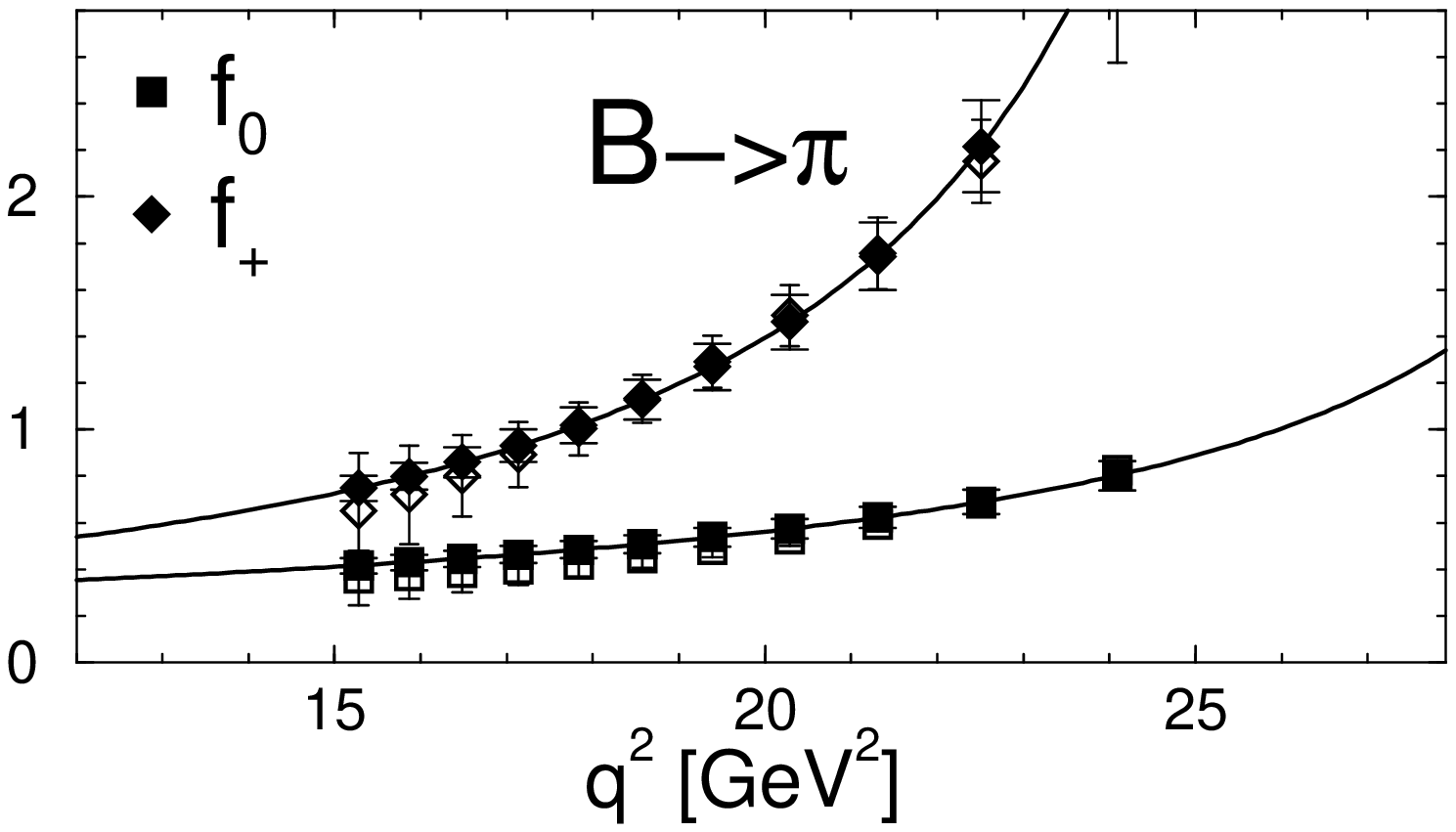,height=37mm}}
\vspace{-.3cm}
\caption{Form factors from 
BK-based (filled symbols and curves) and non-BK-based (open symbols) analyses
for $D\to\pi l\nu$ (left panel) and $B\to\pi l\nu$ (right) decays.}
\label{fig:B2pi}
\vspace{-1cm}
\end{center}
\end{figure}

We first extract the
form factors $f_\parallel$ and $f_\perp$,
and 
interpolate and
extrapolate the results for $f_\parallel$ and $f_\perp$ to common values
of $E$ using 
the parametrization of Becirevic and Kaidalov (BK)
\cite{Becirevic:1999kt}. 
We then perform the chiral extrapolation ($m_l\to m_{ud}$) at each $E$
using the NLO correction
in staggered chiral perturbation theory~\cite{Aubin:2004xd}.
%
Finally we convert 
the results for $f_{\perp}$ and $f_{\parallel}$ at $m_l=m_{ud}$,
to $f_+$ and $f_0$.
The results for $f_+$ and $f_0$ are parameterized with the 
BK form~\cite{Becirevic:1999kt},
\bea\label{eq:BK}
f_+(q^2) = \frac{f_+}{(1-\tilde{q}^2)(1-\alpha\tilde{q}^2)},~~~
f_0(q^2) = \frac{f_+}{1-\tilde{q}^2/\beta},
\eea
where $\tilde{q}^2=q^2/m_{H^{*}}^2$.
We obtain~\cite{Aubin:2004ej,Okamoto:2004xg}
$
f_+^{D\to \pi}=0.64(3),~
\alpha^{D\to  \pi}=0.44(4),~ \beta^{D\to \pi}=1.41(6)$ and
$f_+^{D\to  K}=0.73(3),~  
\alpha^{D\to  K}=0.50(4),~  \beta^{D\to  K}=1.31(7)
$
for the $D$ decays, and 
$
f_+^{B\to\pi}=0.23(2),~ \alpha^{B\to\pi}=0.63(5),~ \beta^{B\to\pi}=1.18(5)
$
for the $B$ decay, where the errors are statistical only. 
To estimate the error from the BK parameterization, {\it i.e.,}
the error for $q^2$ dependence,
we also make an alternative analysis, where we perform a 2-dimensional 
polynomial fit in $\left(m_l,E(q^2)\right).$ 
A comparison between two analyses is shown in Fig.~\ref{fig:B2pi}.
The results for $D$ decays agree well with recent experimental 
results \cite{unknown:2004nn}.
We then determine the CKM matrix elements 
by integrating $|f_+(q^2)|^2$ over $q^2$ and 
using experimental decay rates~\cite{Eidelman:wy,Athar:2003yg,Belle:B2pi}.
For $|V_{ub}|$ we use a combined average of 
the decay rate for $q^2 \ge 16$ GeV$^2$
in Refs.~\cite{Athar:2003yg} and \cite{Belle:B2pi}.
We obtain
\bea
~~~|V_{cd}| = 0.239(10)(24)(20) \,\,\,\,\, ,\,\,\,\,\,
   |V_{cs}| = 0.969(39)(94)(24)
\label{eq:VcdVcs}
\eea
from the $D$ decay, and 
\bea
~~~|V_{ub}|\times 10^{3} = 3.48(29)(38)(47) 
\label{eq:Vub}
\eea
from the $B$ decays, where the first errors are statistical,
the second systematic, and the third are experimental errors from
the decay rates.
The systematic errors 
are dominated by the finite lattice spacing effects,
{\it i.e.,} the lattice discretization effects; 
see Table~\ref{tab:error}.
The results for the CKM matrix elements agree with the 
Particle Data Group averages~\cite{Eidelman:wy} with a comparable accuracy.

\begin{table}[b]
\begin{center}
\caption{Systematic errors in lattice calculations. 
For comparison, the error for each CKM matrix element by the
Particle Data Group~\protect\cite{Eidelman:wy} is shown in the last row.}
{\begin{tabular}{l|rrrr}
\hline
semileptonic decay  &$D\to \pi(K)l\nu$ & $B\to \pi l\nu$ & $B\to D l\nu$ & 
$K\to \pi l\nu $\\
CKM matrix element  &$|V_{cd(s)}|$  &$|V_{ub}|$   &$|V_{cb}|$ & $|V_{us}|$\\
\hline
$q^2$ dependence & 2\% & 4\% &  & $<$1\%\\
$m_l\!\to\! m_{ud}$ extrapolation& 3\%(2\%) &4\% & 1\% & 1\%\\
operator matching & $<$1\% & 1\% &  1\%  & $<$1\%\\
{discretization effects} & { 9\%}& { 9\%} &  $<$1\%  & \\
\hline
total systematic error   & { 10\%} & { 11\%} &  { 2\%}  & {1\%}\\
\hline
\hline
\vspace{-.4cm}\\
error in PDG average
& 5\%(1\%)       & 13\%       &  4\%    & 1\%\\
\hline
\end{tabular}\label{tab:error}}
\end{center}
\end{table}

\subsection{$B\to D l\nu$, $|V_{cb}|$ and $K\to \pi l\nu$, $|V_{us}|$}
%
The differential decay rate of $B\rightarrow D l\nu$ is
proportional to the square of $|V_{cb}| {\cal{F}}(w)$, 
where ${\cal{F}}(w)$ is 
the relevant form factor and 
$w=v\cdot v'$ with $v=p_B/m_B$, $v'=p_D/m_D$.
To extract $|V_{cb}|$,
we calculate the form factor at $w=1$, ${\cal{F}}(1)$,
by employing the double ratio method \cite{Hashimoto:1999yp}.
The light quark mass dependence of ${\cal{F}}(1)$ is 
mild, and by
extrapolating the result linearly to $m_l\to 0$ 
we obtain~\cite{Okamoto:2004xg}
$
{\cal{F}}^{B\rightarrow D}(1)=1.074(18)(16),
$
where the first error is statistical, and the second is systematic.
Combining with the experimental result for 
$|V_{cb}{|\cal{F}}(1)$~\cite{HFAG}, we obtain 
\bea
   |V_{cb}|\times 10^{2} = 3.91(07)(06)(34),
\label{eq:Vcb}
\eea
where the first two errors from lattice calculation are smaller than
the experimental one (third).

\begin{figure}[tb]
\begin{center}
\epsfig{file=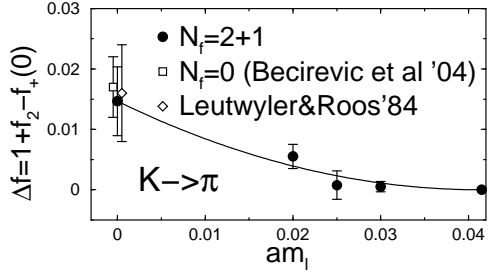,height=40mm}
\vspace{-.5cm}
\caption{$m_l$-dependence of $\Delta f$ for $K\to\pi l\nu$ decay,
together with results from Refs.~\protect\cite{Becirevic:2004ya,Leutwyler:1984je}.}
\label{fig:K2pi}
\vspace{-.6cm}
\end{center}
\end{figure}

Finally we study the $K\to\pi l\nu$ decay to determine $|V_{us}|$.
The expression for the $K\to\pi$ decay amplitude is given in an
analogous way to Eq.~(\ref{eq:HLff}).
We calculate the $K^0\to\pi^-$ form factor at $q^2=0$, $f_+(0)=f_0(0)$,
by employing the three-steps method, as in 
Ref.~\cite{Becirevic:2004ya}.
To perform the $m_l\!\to\! m_{ud}$ extrapolation for $f_+(0)$,
we subtract the leading logarithmic correction $f_2$
in chiral perturbation theory,
{\it i.e.,} define $\Delta f\equiv 1+f_2-f_+(0)$.
We make a fit to $\Delta f$ adopting an ansatz,
$\Delta f = (A +B m_l)(m_s-m_l)^2$, where $A,B$ are fit parameters.
The $m_l$-dependence of $\Delta f$ and the extrapolated result
are shown in Fig.~\ref{fig:K2pi}, together with 
a recent quenched lattice result~\cite{Becirevic:2004ya}
and an earlier result by 
Leutwyler and Roos~\cite{Leutwyler:1984je}.
Our preliminary result~\cite{Okamoto:2004df}
is $\Delta f = 0.015(6)(9)$, giving
$
{f}_{+}^{K^0\to\pi^-}(0) = 0.962(6)(9),
$
which agrees well with those of  
Refs.~\cite{Becirevic:2004ya,Leutwyler:1984je}.
Combining with a recent experimental result for 
$|V_{us}|{f}_{+}(0)$~\cite{Alexopoulos:2004sw},
we obtain
\bea
|V_{us}| = 0.2250(14)(20)(12).
\label{eq:Vus}
\eea

\section{Other 4 CKM matrix elements using unitarity and 
Wolfenstein parameters}\label{sec:unitary}

Having the 5 CKM matrix elements directly determined from
the 5 semileptonic decays, we can check unitarity
of the second row of the CKM matrix. Using 
Eqs.~(\ref{eq:VcdVcs}) and (\ref{eq:Vcb}), we get
\bea
({|V_{cd}|^2+|V_{cs}|^2+|V_{cb}|^2})^{1/2}={ 1.00(10)(2)},
\eea
which is consistent with unitarity.
Hereafter the first error is from the lattice calculation and the second
is experimental, unless otherwise stated.

We now use unitarity of the CKM matrix to determine
the other 4 CKM matrix elements.
$|V_{ud}|$, $|V_{tb}|$ and $|V_{ts}|$ are easily determined:
\bea
|V_{ud}| &=& (1 - |V_{us}|^2 - |V_{ub}|^2)^{1/2} ~=~ \Vud, \\
|V_{tb}| &=& (1 - |V_{ub}|^2 - |V_{cb}|^2)^{1/2} ~=~ \Vtb, \\
|V_{ts}| &=& |V_{us}^{*}V_{ub} + V_{cs}^{*}V_{cb}|\ /\ |V_{tb}|
~\simeq~ |V_{cs}^{*}V_{cb}|\ /\ |V_{tb}| ~=~ \Vts. 
\eea
Eqs.~(\ref{eq:Vus}), (\ref{eq:Vcb}) and (\ref{eq:Vub}) 
give some of the Wolfenstein parameters,
\bea
\lambda &=& |V_{us}| ~~~~~~~~~~=\!\!\!\! \Wlambda\!\!\!\!\!\!, \\
A &=& |V_{cb}|/\lambda^2 ~~~~~~=~ \WA, \\
(\rho^2 + \eta^2)^{1/2} &=& |V_{ub}| / (A\lambda^3) ~=~ \WR. 
\label{eq:WR}
\eea
%
To extract $|V_{td}|$ and $(\rho,\eta)$,
we use the experimental result for $\sin(2\beta)$
from $B\to (c\bar{c}) K^{(*)}$ decays. 
From a unitary triangle analysis with 
$\sin(2\beta)=0.726(37)$~\cite{HFAG}
and Eq.~(\ref{eq:WR}),
we obtain
\bea
\rho ~=~ \Wrho, &&~ \eta ~=~ \Weta, \\
|V_{td}| &=& \Vtd    
\eea
with a combined lattice and experimental error,
completing the {full} CKM matrix. 

\section*{Acknowledgments}

We thank 
all members of the Fermilab Lattice, MILC and HPQCD Collaborations,
in particular, 
the authors in Ref.~\cite{Aubin:2004ej}.
%
This work is supported by
the Fermilab Computing Division, the SciDAC program, and
FermiQCD~\cite{DiPierro:2003sz}.
Fermilab is operated by Universities Research Association Inc., under
contract with the U.S. Department of Energy.

\end{document}